\documentclass{elsart}
\usepackage[dvips]{epsfig}
\begin{document}
 \def \figwidth  {6.0}
\vspace*{-0.6in}
\begin{flushright}
FERMILAB-PUB-02/304-E\\
CDF/DOC/CDF/PUBLIC/5936\\
nuhep-exp/02-014\\
\today\\
\end{flushright}
\vspace{0.5in}

\begin{frontmatter}

 \title{Beam Halo Monitoring at CDF}
 \author[Northwestern]{Muge Karagoz Unel}
 \ead{karagozm@fnal.gov},
 \author[Fermilab]{Richard J. Tesarek}
 \ead{tesarek@fnal.gov}

 \address[Northwestern]{Northwestern University, Evanston, IL 60208, USA}
 \address[Fermilab]{Fermilab, Batavia, IL 60510, USA}

 \begin{abstract}
  Losses from the proton and antiproton beams of the Fermilab Tevatron have 
  been shown to produce a halo which contribute to backgrounds to physics 
  signals in the Collider Detector at Fermilab (CDF). To provide a measure 
  of the beam halo and losses, we have installed arrays of scintillation 
  counters on both sides of the CDF detector. We describe here the physical 
  configuration of these counters, their implementation and performance  
  within the Fermilab Accelerator Control Network (ACNET). We 
  show correlations between these new devices and the accelerator 
  operating conditions. We point out that the use of these monitors 
  is leading to improvement in the accelerator operations and reduced 
  backgrounds in CDF.
 \end{abstract}

\end{frontmatter}

\section{Introduction}
\begin{figure}[ptb]
 \begin{center}
  \epsfxsize=\figwidth in
  \epsffile{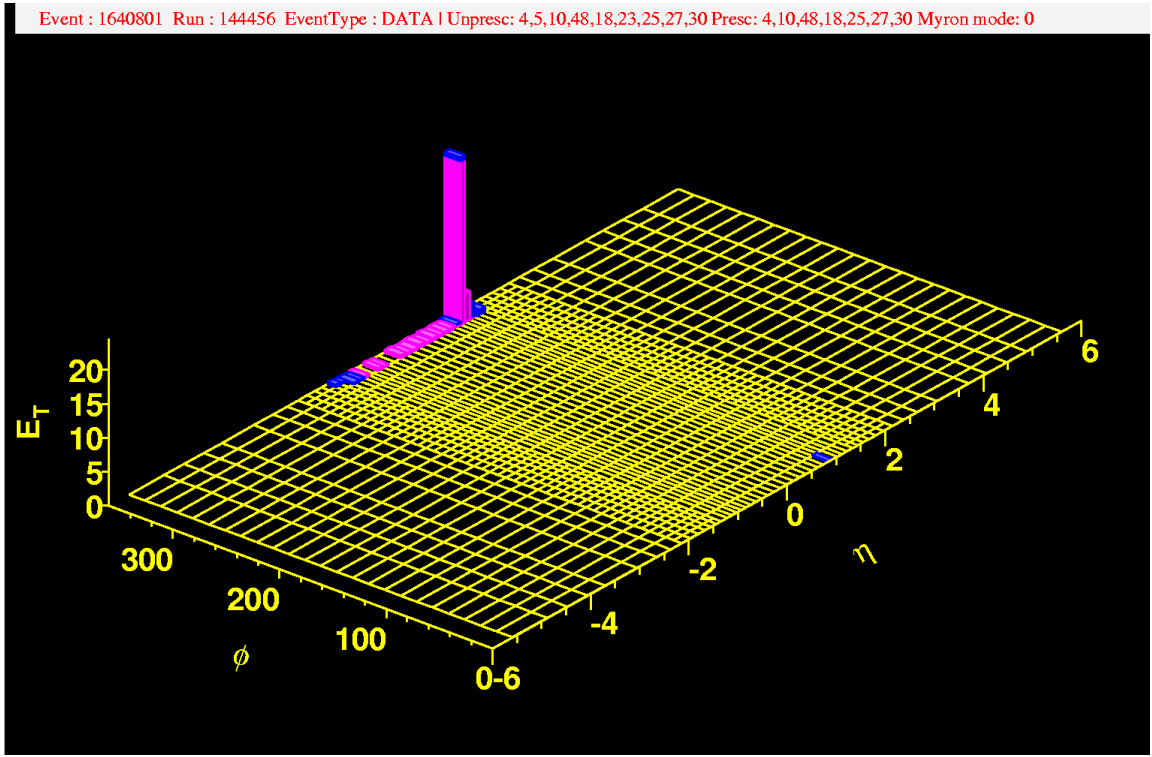}  
    \caption{\label{lego} 
              CDF event display showing the energy deposited in 
              the calorimeters as a function of location in the 
              calorimeter ($\eta,\phi$), where 
              $\eta = -\log({\tan{\frac{\theta}{2}}})$ and $\phi$ 
              is the azimuthal angle.  Protons enter from the 
              $-\eta$ side.
                 }
 \end{center}
\end{figure}

The Run~II program for the Tevatron started in 2001 and the CDF experiment has 
been taking data with recent luminosities comparable with the peak 
luminosity obtained at the end of Run I \cite{runII}. However, the CDF 
detector has encountered operational problems and observed high backgrounds 
which have been correlated with proton and antiproton losses from the 
Tevatron.  For example, high loss rates have caused high anode currents
and consequent chamber trips in most muon detectors, especially in central 
muon upgrade and extension (CMP, CMX) chambers~\cite{muon}.
A number of low voltage switching power supply failures 
were also observed in periods of high beam loss rates associated with 
Tevatron injection and aborts.~\footnote{These power supply failures
were traced to a catastrophic, single event failure (single event burn out) 
in a radiation soft, power MOSFET used in the supplies.} In addition, 
backgrounds in the CDF detector have been associated with a halo of 
particles accompanying the beam.
Figure \ref{lego} shows an event display of an off-axis muon producing a
hard bremsstrahlung in the central calorimeter.  The muon ``track''
is identified by the set of contiguous towers in
   $\eta = -\log{ (\tan{ \frac{\theta}{2} } )}$
at constant azimuth ($\phi$).  The vertical scale is the energy in the
calorimeter transverse to the beam.  These events form backgrounds
to physics signals involving photons and missing transverse energy.

CDF measures beam losses using the beam shower counters 
 located closest to the CDF detector (BSC-1).  The BSC-1 are small 
scintillation counters at a radius between 4--7~cm relative to the 
center of the beam pipe and located approximately 5~m on either side of 
the CDF interaction point (IP).  Losses are calculated as the coincidence 
between the counter signal and beam particles passing the plane of the BSC-1.  
A detailed description of the BSC-1 and the loss calculation may be 
found elsewhere~\cite{bsc}. Because the BSC-1 are small, located near 
the beam line and near to the CDF IP, they are ideal for monitoring 
the small angle beam losses which affect various components in the 
tracking volume.  However, these counters do a poor job of measuring 
the beam halo affecting the outer regions of the detector.  Further, the 
raw BSC-1 photomultiplier tube (PMT) signals are unavailable due to the 
electronics requirements for the system.  These PMTs have also shown 
strong afterpulsing~\cite{binkley_vidal}. Both of these issues make 
additional timing coincidences problematic.

We address these concerns by installing a set of counters (halo monitors)
in the CDF collision hall.  In this article, we describe the physical 
configuration of the halo monitors and their readout. These monitors 
detect the halo particles at a radial distance of about 50~cm around 
the beam line in full azimuthal coverage and at maximum available distance 
from the CDF detector. The monitors are capable of detecting particles 
at larger angles from the beam line and are complementary to the beam 
shower counters.

\section{Setup}
\begin{figure}[ptb]
 \begin{center}
  \epsfxsize=\figwidth in
  \epsffile{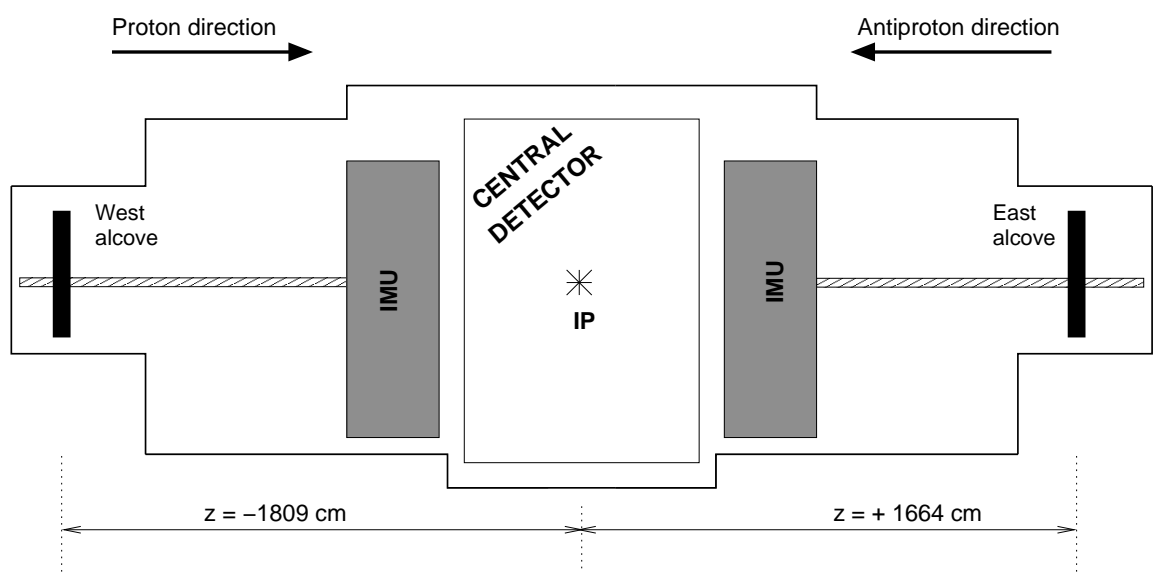} 
  \caption{\label{collhall} 
            Elevation of the CDF collision 
            hall showing the positions of the halo monitors (not 
            to scale). The proton monitor is located at $z = -1809$~cm 
            and the antiproton monitor is at $z = + 1664$~cm.
            }
 \end{center}
\end{figure}

\begin{figure}[ptb]
 \begin{center}
  \epsfxsize=\figwidth in
  \epsffile{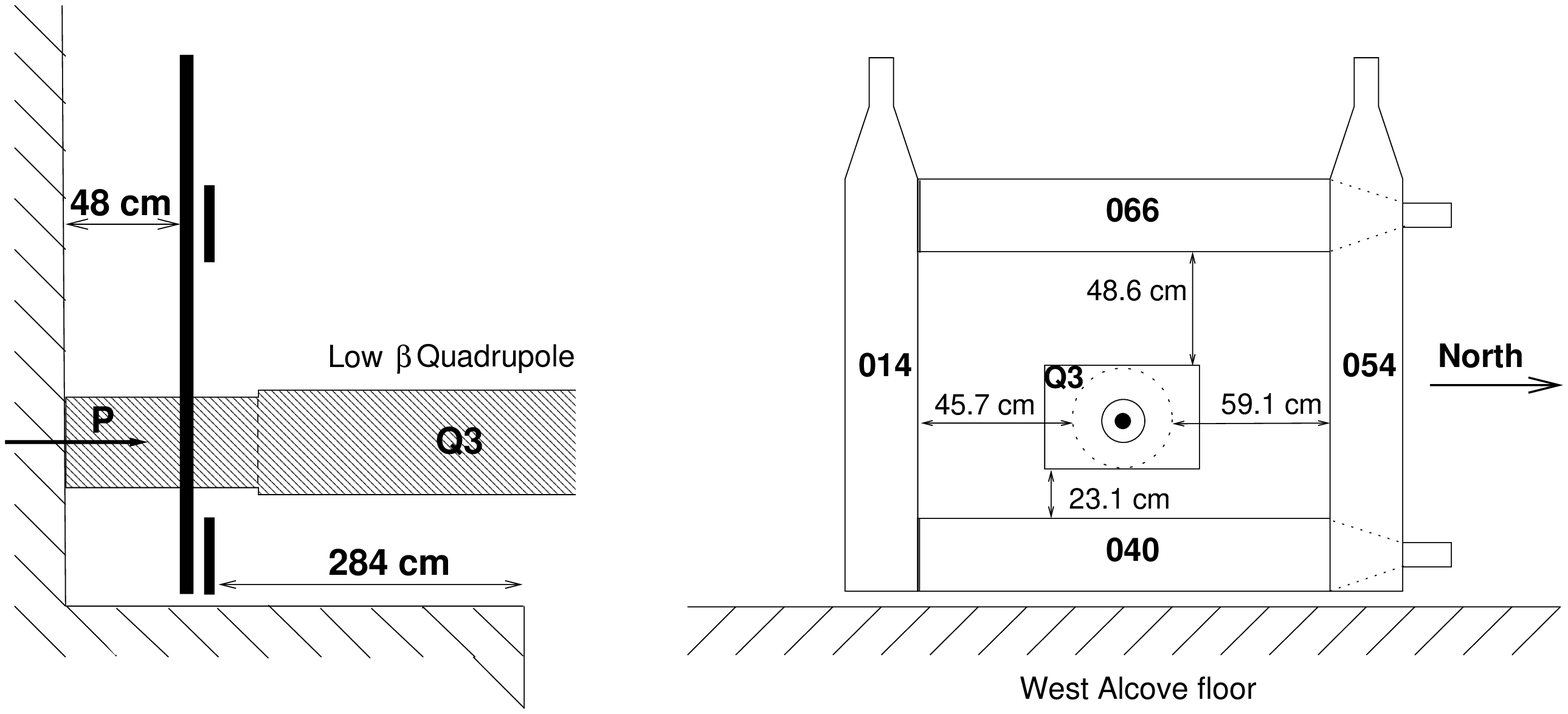}
  \caption{\label{pr_conf} 
            Configuration of the proton halo counters around 
            the low $\beta$ quadrupole in the west alcove. The left 
            figure is the side view and the right figure shows the 
            view with protons exiting out of the page.
            } 
  \vspace{15pt}
 \end{center}
\end{figure}

\begin{figure}[ptb]
 \begin{center}
  \epsfxsize=\figwidth in
  \epsffile{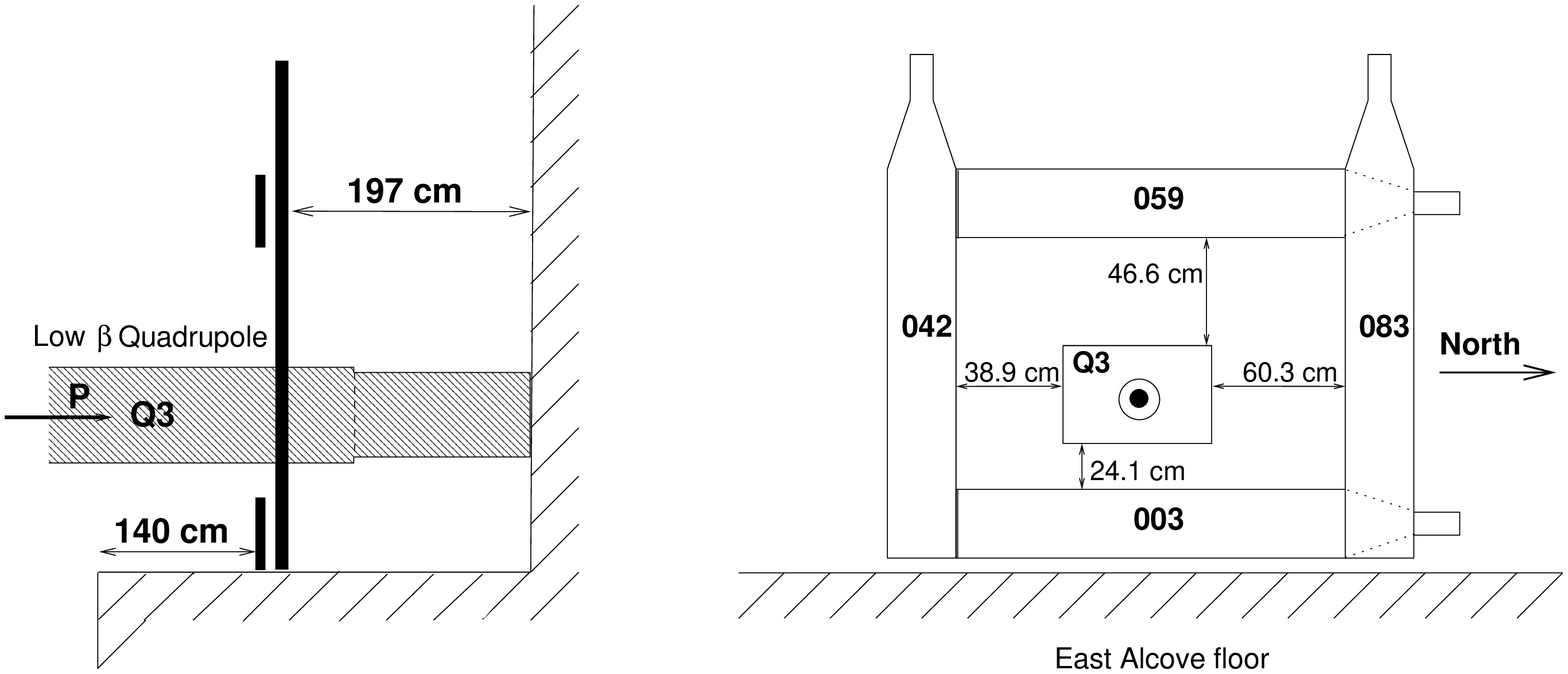}
  \caption{\label{antipr_conf} 
            Configuration of the antiproton halo counters around 
            the low $\beta$ quadrupole in the east alcove. The left 
            figure is the side view and the right figure shows the 
            view with protons exiting out of the page.
            } 
\end{center}
\end{figure}

The CDF detector is located in the B0 section of the Tevatron at 
Fermilab. Two halo monitors  are located in the west and east alcoves 
in the CDF collision hall, at positions $z = - 1809 \rm\; cm$ and  
$z = 1664 \rm\; cm$ with respect to the IP. Both monitors consist 
of an array of four scintillator counters arranged in a non-overlapping 
rectangle surrounding the  final focus, low $\beta$ quadrupoles. 
Figures~\ref{collhall}--\ref{antipr_conf} illustrate the layout of 
the monitors. 

\begin{figure}[ptb]
 \begin{center}
  \epsfxsize=\figwidth in
  \epsffile{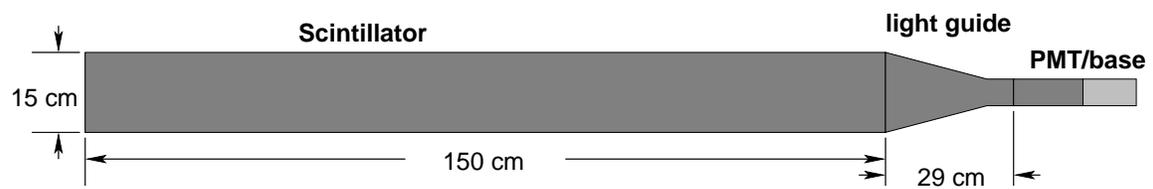} 
  \caption{\label{paddle} 
            Diagram of one scintillator counter 
            used  for the halo monitors.
            }
 \end{center}
\end{figure}

The counters used for these monitors come from a set of spares
originally manufactured for the KTeV muon identification and
veto systems~\cite{ktev}.  These counters are made of Bicron BC-412 
scintillator \cite{bicron_scint} and have an active volume of
$1.3\times 15\times 150 \rm\; cm^3$.
The scintillator is glued to a wedge shaped, lucite light guide
which is in turn glued to an EMI-9954KB PMT~\cite{emi_pmt}. 
Bicron, BC-600, optical cement is used for all glue joints. The end of the
scintillator opposite the PMT was painted black to avoid 
signals from light reflected off the far end. The
counter assemblies are then wrapped in aluminum foil and 76.2~$\mu$m
thick black plastic.  The wrapping materials are held in place
using black vinyl tape.  Figure~\ref{paddle} is a diagram of a single counter. 

Each counter is characterized, after assembly, using three 
cosmic ray telescopes.  The three telescopes identify cosmic 
rays passing through each end and the middle of the counters. 
Operating characteristics including
the relative gain, dark rate and counter efficiency 
are recorded for each telescope as a function of PMT bias voltage  
with a discriminator threshold of 30~mV. 
The operating voltage for 
each counter is chosen to be 100~V above the ``knee'' of the efficiency 
vs PMT voltage (plateau) curve.  The knee is defined to be the 90\% efficiency
point on the plateau curve as measured with the telescope farthest from the 
PMT.  The light attenuation length of each counter is measured at the 
operating voltage.

A high voltage power supply (Power Designs model HV-1547) provides 
bias voltages for all the counter PMTs. The individual values are 
set by a Berkeley high voltage zener divider box.  After installation, 
the ungated singles rate (dark current) for each counter is measured 
at the same discriminator settings and PMT bias voltages used during
the counter production tests described above. Table~\ref{scintillators} 
summarizes the operating parameters for these counters. 

\begin{figure}[ptb]
 \begin{center}
  \epsfxsize=\figwidth in
  \epsffile{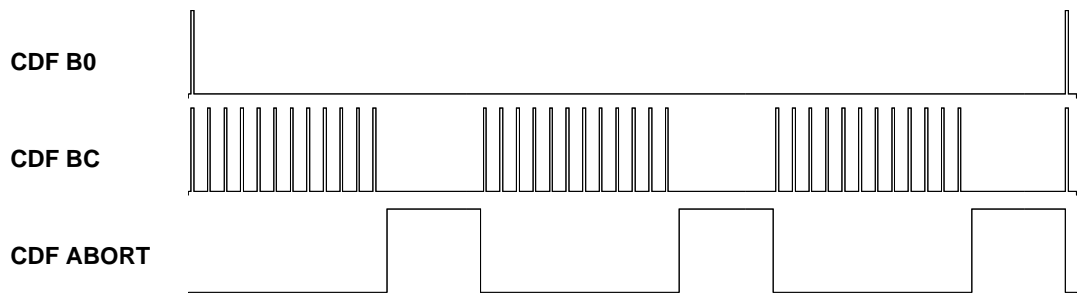}
  \caption{\label{cdftiming} 
            Accelerator timing signals. The time between successive 
            CDF\_B0 events is one Tevatron revolution (20.7~$\mu$s).
            }
 \end{center}
\end{figure}

Beam halo is measured by forming a coincidence between the counter 
signals and a beam bunch signal as the bunch passes through the plane 
of the counter array on its way to the CDF IP. 
To mark the beam crossings, we use the 38~ns wide Bunch Crossing
(CDF\_BC) signal.  The CDF\_BC signal defines the CDF collision
times with a repetition period of 396~ns.   CDF\_BC is derived from 
the 53~MHz Tevatron RF and is received from a  ``TRigger And 
Clock + Event Readout'' (TRACER) module~\cite{tracer}. The TRACER 
module also distributes the timing signals which mark the first 
bunch in a revolution (CDF\_B0) and CDF\_ABORT which mark the 
``abort gaps'' where there are no bunch crossings. All such accelerator 
timing signals are fanned out to TRACER modules. Figure \ref{cdftiming} 
shows the relevant clock signals.

\begin{figure}[ptb]
 \begin{center}
  \epsfxsize=\figwidth in
  \epsffile{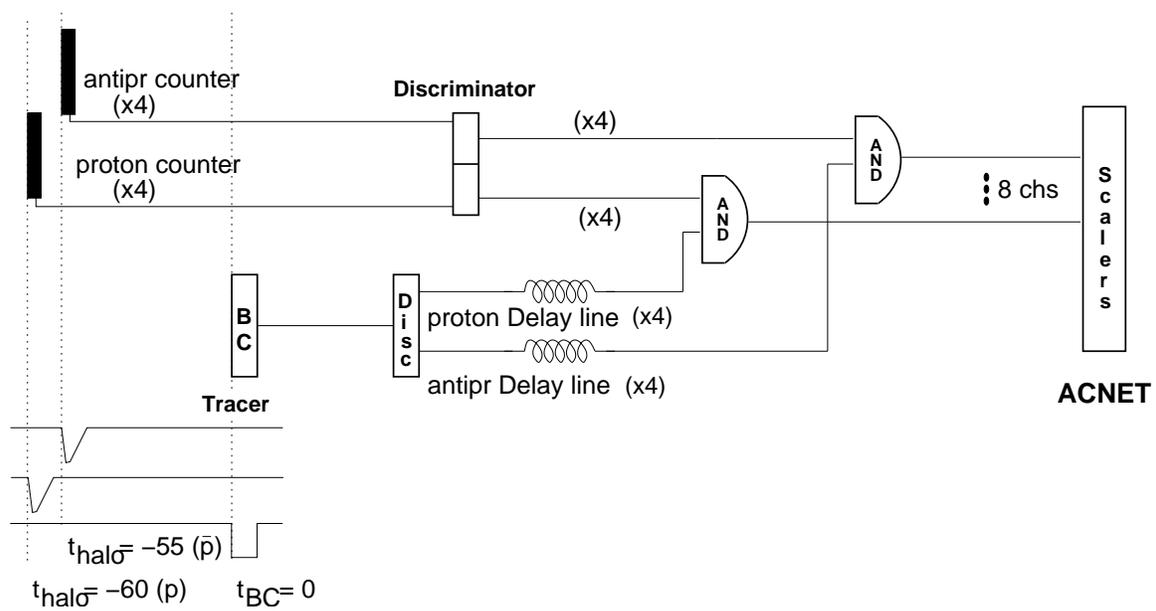}
  \caption{\label{logic_setup}
            Logic diagram for the beam halo 
            monitors. The relative timing is also shown.
            }
 \end{center}
\end{figure}

\begin{figure}[ptb]
 \begin{center}
  \epsfxsize=\figwidth in
  \epsffile{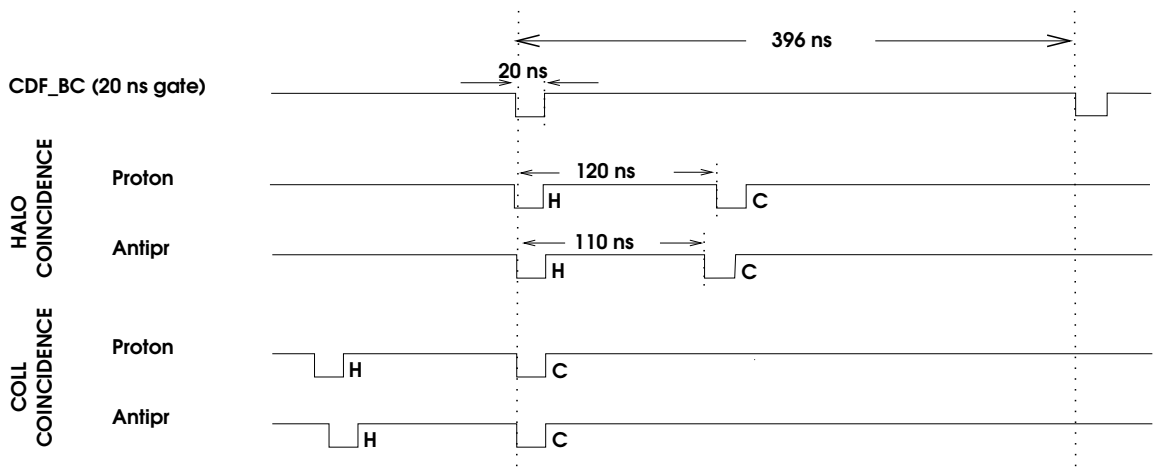}
  \caption{\label{timing_setup} 
            Timing diagram for halo coincidences.  It is possible to 
            trigger for both halo (H) and collision debris (C) by 
            application of appropriate delays.
            }
 \end{center}
\end{figure}

The counter PMT signals are brought up  to a NIM crate in a CDF 
counting room  where they are discriminated using 
a LeCroy~4413 discriminator (threshold $30 \rm\; mV$, width $20 \rm\; ns$). 
The discriminator width is set to allow for light propagation 
delays (8~ns) and slewing for the counters.  The discriminated counter 
signals are carried to a Programmable Logic Unit (PLU) 
(LeCroy model 4516) to form the coincidence with the CDF\_BC.
The CDF\_BC signal, originally of TTL-standards, is first converted to NIM 
standard  and then its width is decreased to $20 \rm\; ns$ by a 
discriminator (LeCroy model 621L). The resulting signal is 
time-aligned with the counter signals using a delay line. The 
time-alignment accounts for the beam particle time-of-flight 
(TOF) difference between the plane of the arrays and the interaction 
time.  A Fan Input/Output Unit is utilized to obtain four copies of 
this signal for the proton monitor counters. Before making another 
four copies for the antiproton counters, an additional 5~ns delay 
is applied to account for TOF difference between the proton and the antiproton 
side.  All CDF\_BC duplicates are further passed through a NIM-ECL 
converter to be sent to the PLU. The eight outputs from the PLU are 
counted by CAMAC scalers.
Figure~\ref{logic_setup} shows the logic diagram for the readout.  
The time alignment of the discriminated signals at the {\sc AND} 
gate is shown in Figure~\ref{timing_setup}.

\begin{figure}[ptb]
 \begin{center}
  \epsfxsize=\figwidth in
  \epsffile{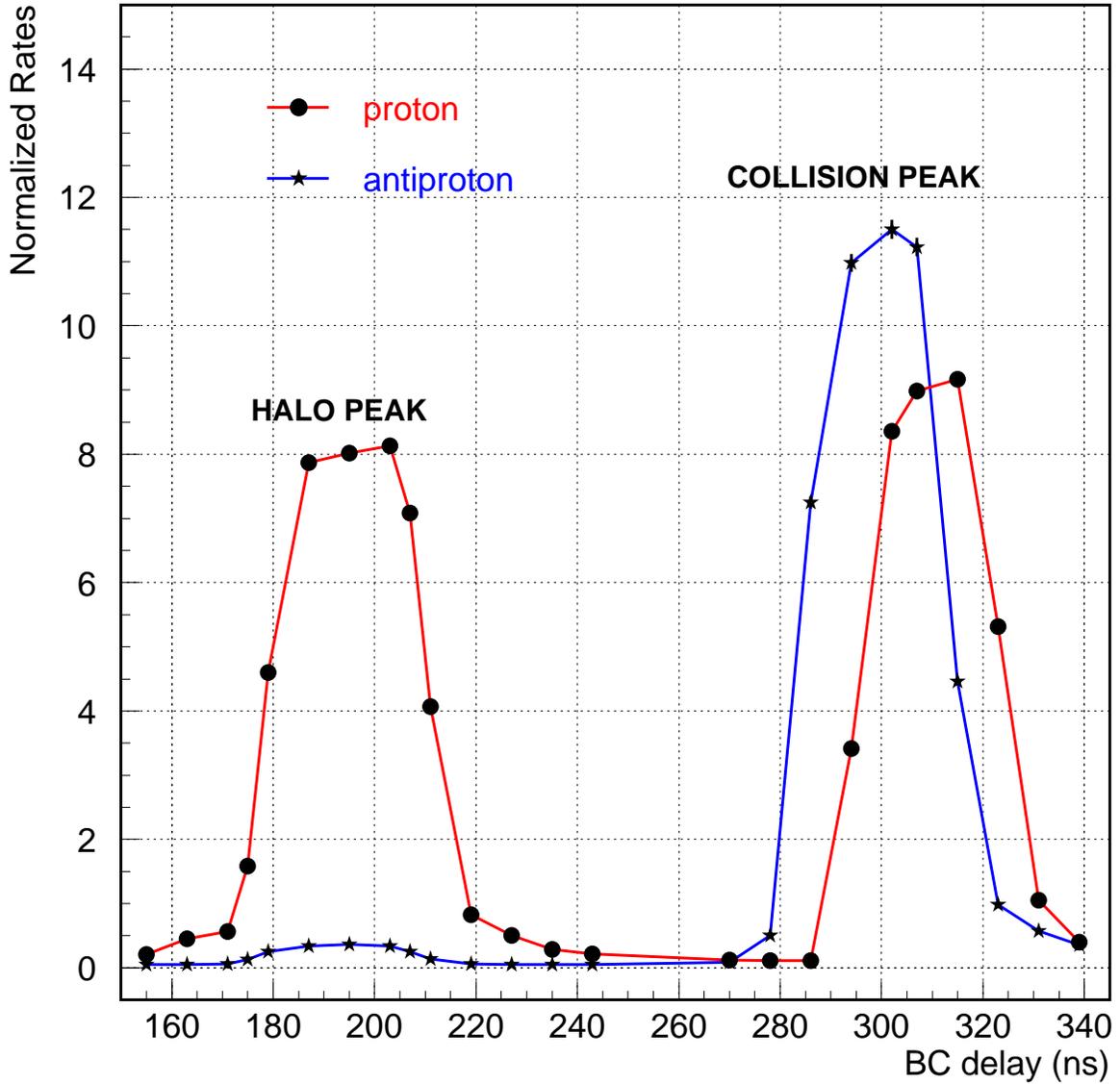}
  \caption{\label{delay_curve1}
            Halo counter rates vs CDF\_BC delay. The left peak is
            from the incoming beam halo; the right peak is from collisions. 
            Statistical errors are also shown at each data point. The 
            differing size of the halo peaks is due to the 
            difference in the proton and antiproton beam currents.
            }
  \vspace{15pt}
 \end{center}
\end{figure}

\begin{figure}[ptb]
 \begin{center}
  \epsfxsize=\figwidth in
  \epsffile{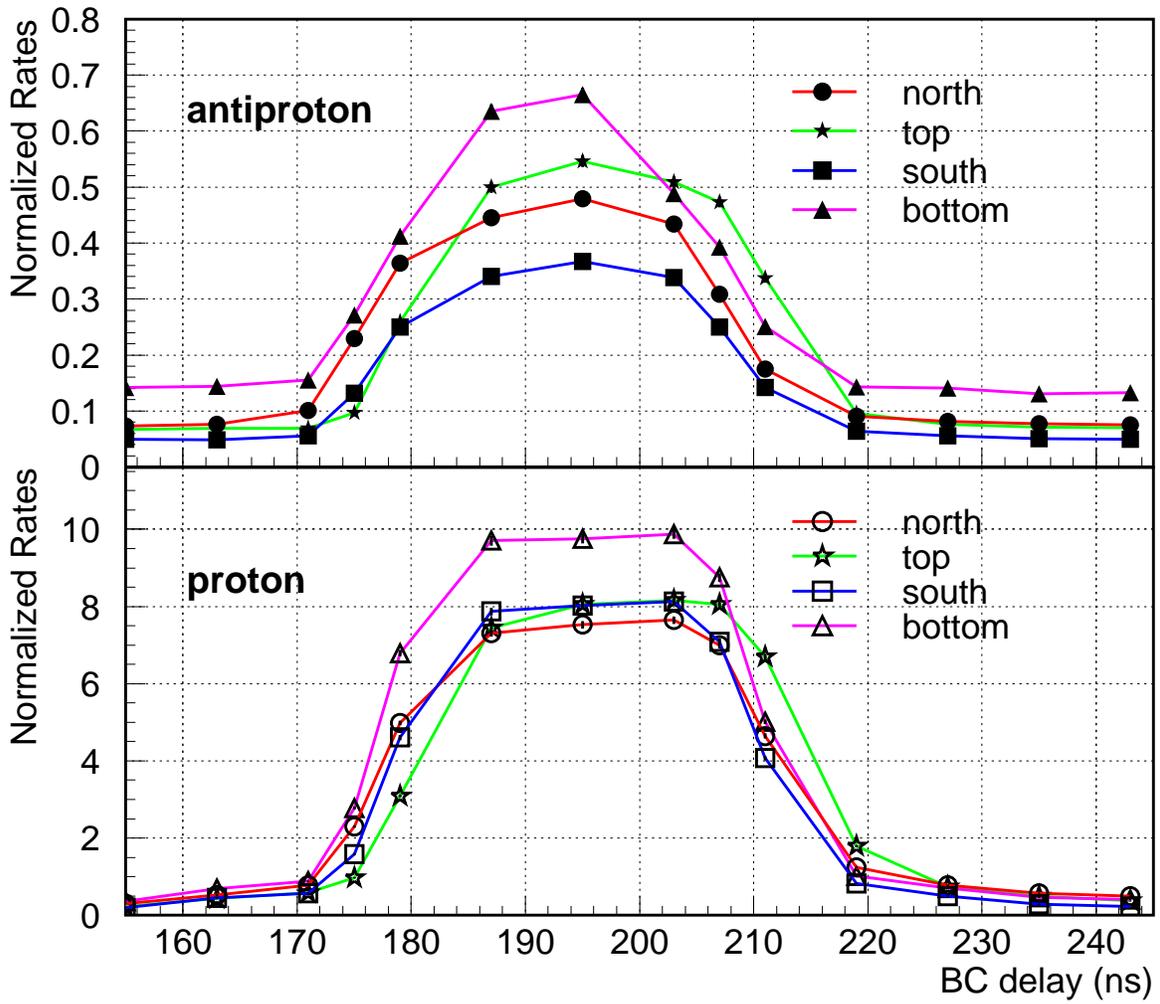}
  \caption{\label{delay_curve2} 
            The portion of the CDF\_BC delay curve 
            showing the halo coincidence plateaus.
}
 \end{center}
\end{figure}

The delay setting for halo coincidences is determined using a 
delay (coincidence) curve technique. Given a fixed counter signal 
propagation time, coincidence rates are measured as a function of 
the delay applied to the CDF\_BC signal. We estimate the CDF\_BC delay necessary
for the coincidence by measuring the cable lengths, module propagation delays and calculating
the flight delays. We then installed a delay cable 
approximately 40~ns shorter than this expected length and measured the 
coincidence rates incrementing the delay by 8~ns between measurements.  
The coincidence data is read out by visual scalers and rates are 
calculated by sampling for 10 seconds and dividing the counts by 10. 
Effects due to beam intensity variations in time are taken into account
by simultaneously reading from a separate, ungated counter.  The decrease
observed in the rates read by the ungated counter at the end of the data taking 
is 6\%. 
The coincidence rates from the counters are normalized to the rate 
measured with the ungated counter and delay curves are obtained using 
these normalized rates. The halo CDF\_BC delay is chosen to correspond 
to the mid-points of the coincidence plateaus for all of the eight 
counters. Figure~\ref{delay_curve1} shows the delay curves for all 
counters. The figure exhibits a collision peak as well as a halo 
peak. The collisions follow the halo signals at twice the particle 
TOF between the monitors and the IP. The difference of about 10~ns 
between the collisions for protons and antiprotons is due to  
the different $z$ positions of the counter arrays.   
The ratio of proton to antiproton beam currents of approximately 10:1 is 
clearly reflected in the halo peaks.
Figure~\ref{delay_curve2} details the halo coincidence curves for 
each counter. The full width at half maximum of the coincidences 
is $\sim$~35~ns as expected.

The halo coincidence signals are read out through two CAMAC 
scalers (Fermilab ``Beams Division'' model~333~\cite{333}). 
These scalers are a part of the Fermilab Accelerator Controls Network 
(ACNET). ACNET is a system to monitor 
and control the Fermilab accelerator complex. A graphical interface  
gives one access to accelerator data in real time and previous data from archives.
The rates for the individual counters 
are calculated \footnote{This calculation is implemented within the 
ACNET software framework.} using the expression: 
\begin{equation}
 \label{counter_eq}
 \hbox{Rate}=\frac{N_{2,i}-N_{1,i}}{N_{2,0}-N_{1,0}}\times 
 \hbox{(Rate of channel 0),}
\end{equation}
where $N_{1,i}$ and $N_{2,i}$ are the two consecutive readings of scaler 
channel $i$ which corresponds to an ACNET device. 
Channel~0 of both CAMAC scalers counts the Tevatron beam revolutions (period of  $20.7 \rm\; \mu s$). 

We also calculate the total rate of each counter array within ACNET. 
These summed rates, Proton Halo SuM (C:B0PHSM) and Antiproton Halo 
SuM (C:B0AHSM), are calculated as:
\begin{equation}
 \label{sum_eq}
 \hbox{Rate}=\frac{\sum{(N_{2,i}-N_{1,i})}}{N_{2,0}-N_{1,0}}\times 
 \hbox{(Rate of channel 0)}
\end{equation}
where the sum is over channels, $i$.  

The halo monitor counters are also used to measure the ``DC beam'' 
(component of the beam not captured in bunches) by measuring the 
halo in the abort gaps of one Tevatron cycle. We 
follow the same logic and use the exact 
halo measurement setup except for two details: in addition to the bunch 
crossing signal from the TRACER module as described above, we carry 
an abort gap signal (CDF\_ABORT) to the PLU and form the coincidence 
with the counters. In order to avoid contamination from the collision 
of the last bunch before the abort gap, 116~ns is removed from the leading edge 
of the duplicate CDF\_ABORT gate. The abort gap coincidence signals 
correspond to eight scaler channels and their rates are calculated  
using formulae \ref{counter_eq} and \ref{sum_eq}.

The rates calculated as in Equations \ref{counter_eq} and \ref{sum_eq} 
have been available on ACNET since April, 2002. 
The ACNET data for the proton and antiproton  monitor total rates (C:B0**SM) are archived with a minimum update frequency of 1~Hz. The minimum data logging 
rate for individual counters is 1~minute in the CDF logger.

We verify the rate calculations by measuring the singles rates from ACNET
and comparing these rates with those expected from previous measurements.
 We first measured the beam off coincidence rates 
 using ACNET data and calculated the expected values 
given the singles rates from Table \ref{scintillators} as a cross-check. 
For the calculation, we assumed a uniform, 35~ns resolving time for 
 halo coincidences and approximately 2.4~$\mu$s for abort gap 
coincidences. Table~\ref{nobeam} shows a comparison between the 
calculated and measured rates. The quoted uncertainties include 
only the statistical errors. The measured and calculated values 
are in good agreement.

The choice of discriminator widths for the counters, CDF\_BC and CDF\_ABORT 
gates imposes an upper limit on the rates these devices 
can effectively measure.  The maximum (saturation) rate for the 
beam halo measurements is 1.7~MHz (maximum CDF\_BC rate).  The 
summed rates saturate at 4 times the above rate (6.8~MHz).  The 
abort gap rates saturate at 4.6~MHz, limited by the discriminator 
width (20~ns), the retriggering time of the discriminator (2~ns) and  
duty factor of the CDF\_ABORT signal after adjustment. The 
maximum rate of the summed signals is 18.4~MHz.

\section{Performance of the Monitors}

\begin{figure}[tbp]
 \begin{center}
  \epsfxsize=\figwidth in
  \epsffile{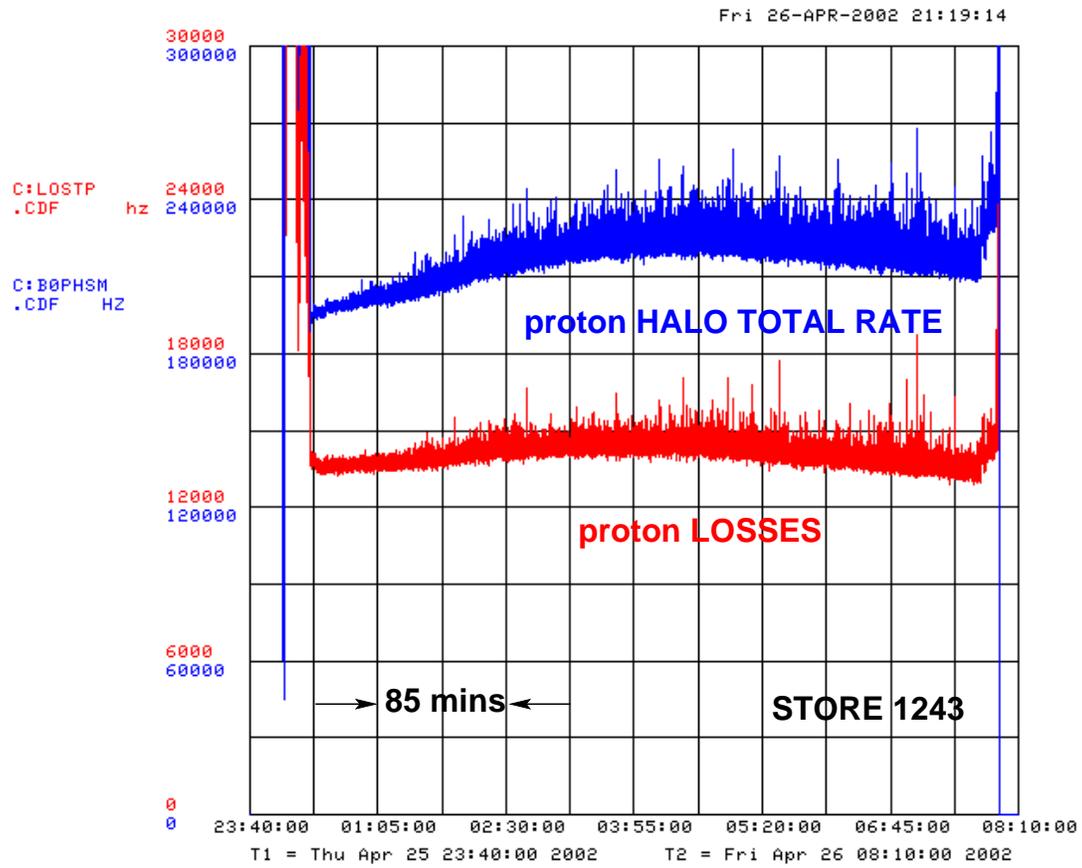}
  \caption{\label{pcompare}
            Total proton halo rates (C:B0PHSM, bottom scale values)  
            and proton losses (C:LOSTP, top scale values)  
            vs time for store~1243.
            } 
 \end{center}
\end{figure}

\begin{figure}[tbp]
 \begin{center}
  \epsfxsize=\figwidth in
  \epsffile{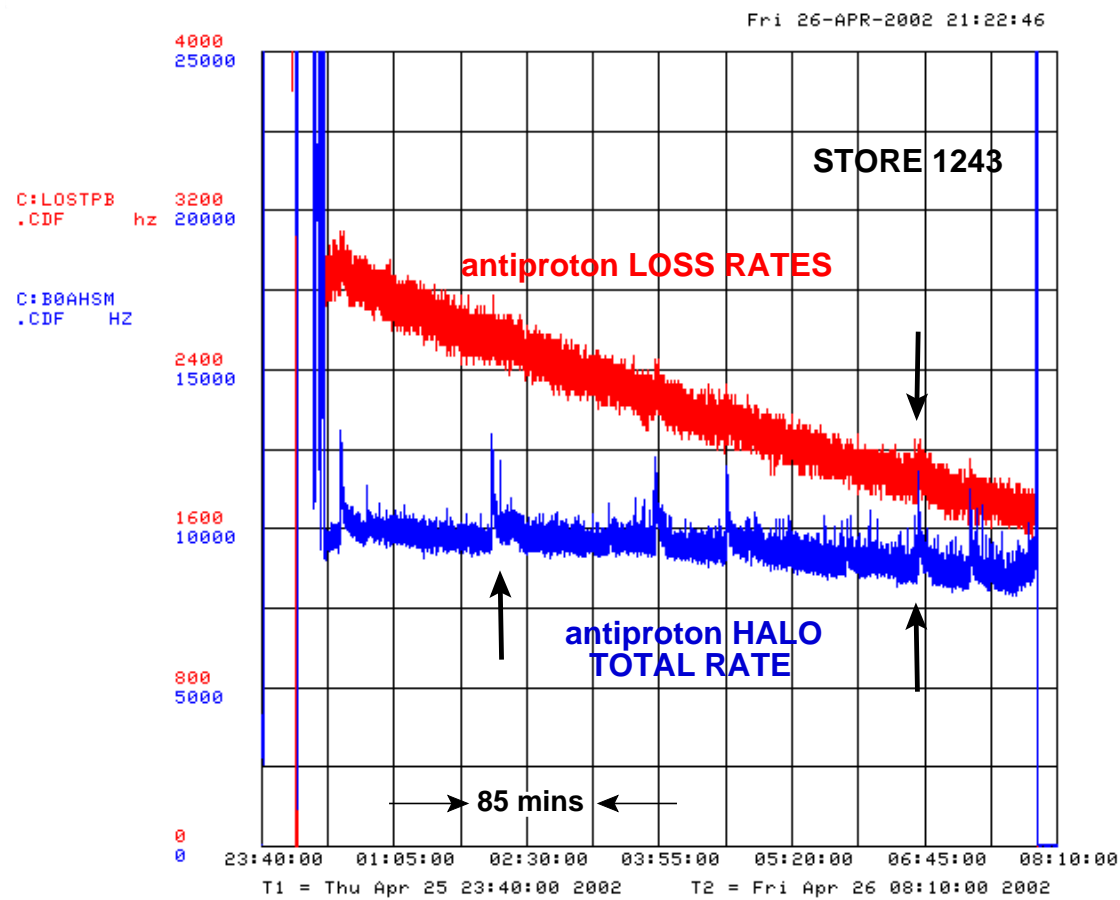}
  \caption{\label{antipcompare}
            Total antiproton halo rates (C:B0AHSM, bottom scale values) 
            and loss rates (C:LOSTPB, top scale values) vs 
            time for store~1243. Arrows indicate new features previously 
            unnoticed.
            }
 \end{center}
\end{figure}

A first look at the performance of the monitors shows that they 
are indeed sensitive to the proton and antiproton losses. 
Figures~\ref{pcompare} and~\ref{antipcompare} compare  halo monitor 
rates with the beam losses measured at CDF. These data are taken 
during Tevatron store~1243. Note the clear features in the antiproton halo 
rates that are just visible in the antiproton loss rates 
(Figure~\ref{antipcompare}). These features exist only in the 
antiproton monitors and are to be examined in more detail.

\begin{figure}[tbp]
 \begin{center}
  \epsfxsize=\figwidth in
  \epsffile{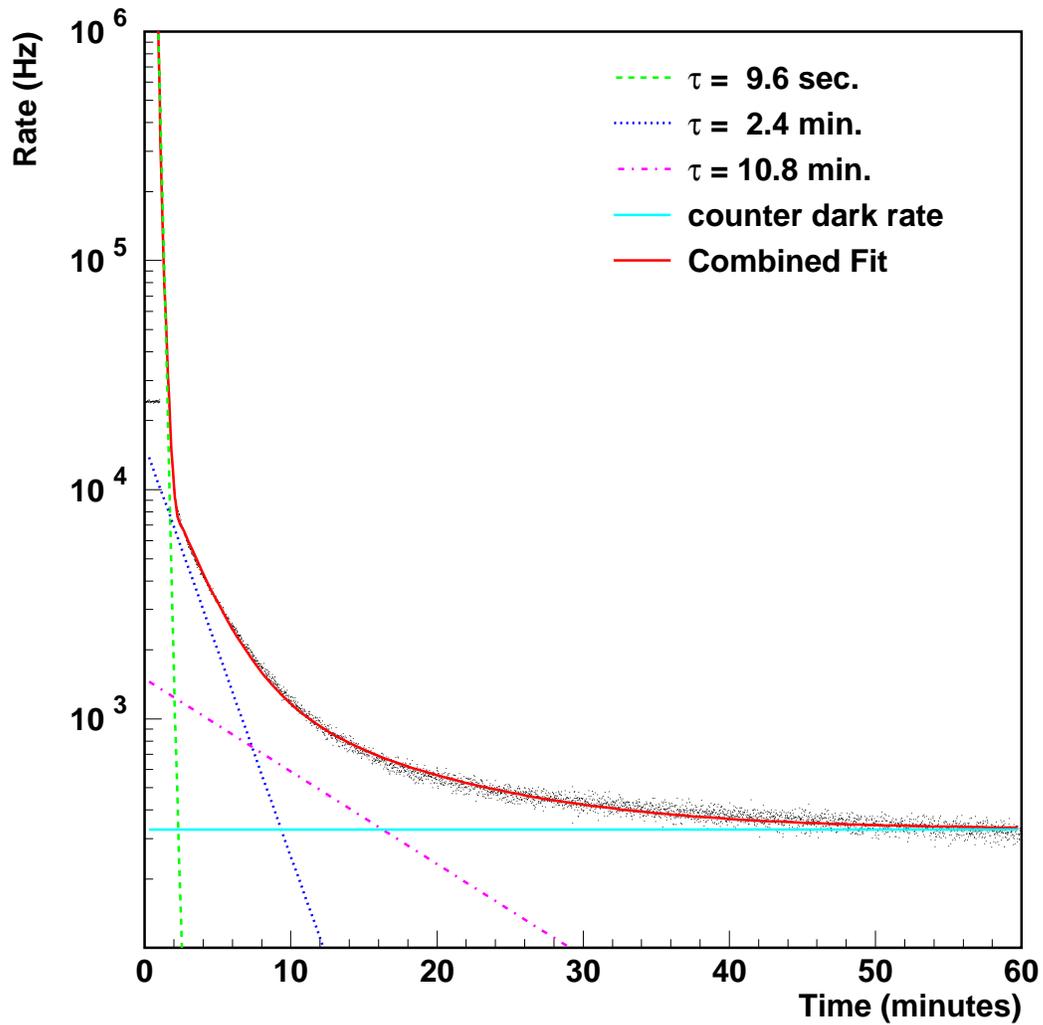}
  \caption{\label{potassium}
            Proton abort gap halo rate vs time after a Tevatron beam 
            abort. The curve represent a fit to three exponentials 
            plus a constant.  The decay constants are consistent with 
            the decays from three short lived potassium isotopes: 
            $^{38}$K, $^{46}$K and $^{48}$K.
            }
 \end{center}
\end{figure}

While commissioning the abort gap halo monitors, we found that
the wide gate allows a significant accidental contamination
from beam induced radioactivity.  This
is best illustrated in Figure~\ref{potassium} where the beam was 
aborted and the abort gap rates did not immediately go to their cosmic
ray value. Performing a lifetime analysis of the above incident, one 
finds that three, short lived isotopes of potassium are responsible
for the increase of the abort gap halo rate. This increased rate is 
removed by requiring a 2/4 majority in each abort gap halo measurement. 
The majority coincidence rates form two ACNET devices which are 
C:B0PAGC (proton) and C:B0AAGC (antiproton). The data logging and 
saturation characteristics of these devices are the same as the 
previously listed abort gap halo monitor devices.

\begin{figure}[tbp]
\begin{center}
  \epsfxsize=\figwidth in
  \epsffile{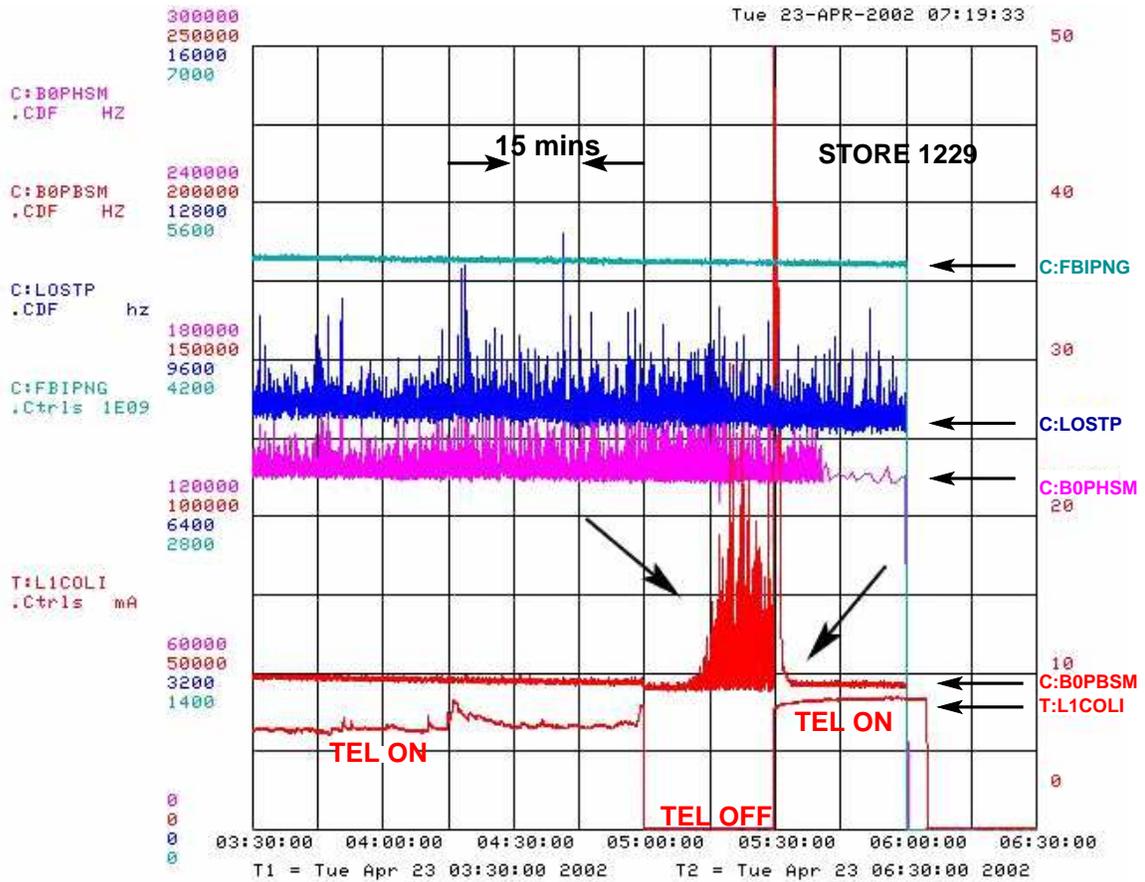}
  \caption{\label{pabort}
            Various ACNET variables vs time for a portion of store~1229. 
            The large arrows indicate the sensitivity of proton 
            abort gap halo monitor (C:B0PBSM) to Tevatron electron lens 
            (T:L1COLI) being turned on. C:FBIPNG is the proton beam current 
            in the Tevatron. The TEL was off for approximately half an hour.
            }
  \vspace{15pt}
 \end{center}
\end{figure}

\begin{figure}[ptb]
 \begin{center}
  \epsfxsize=\figwidth in
  \epsffile{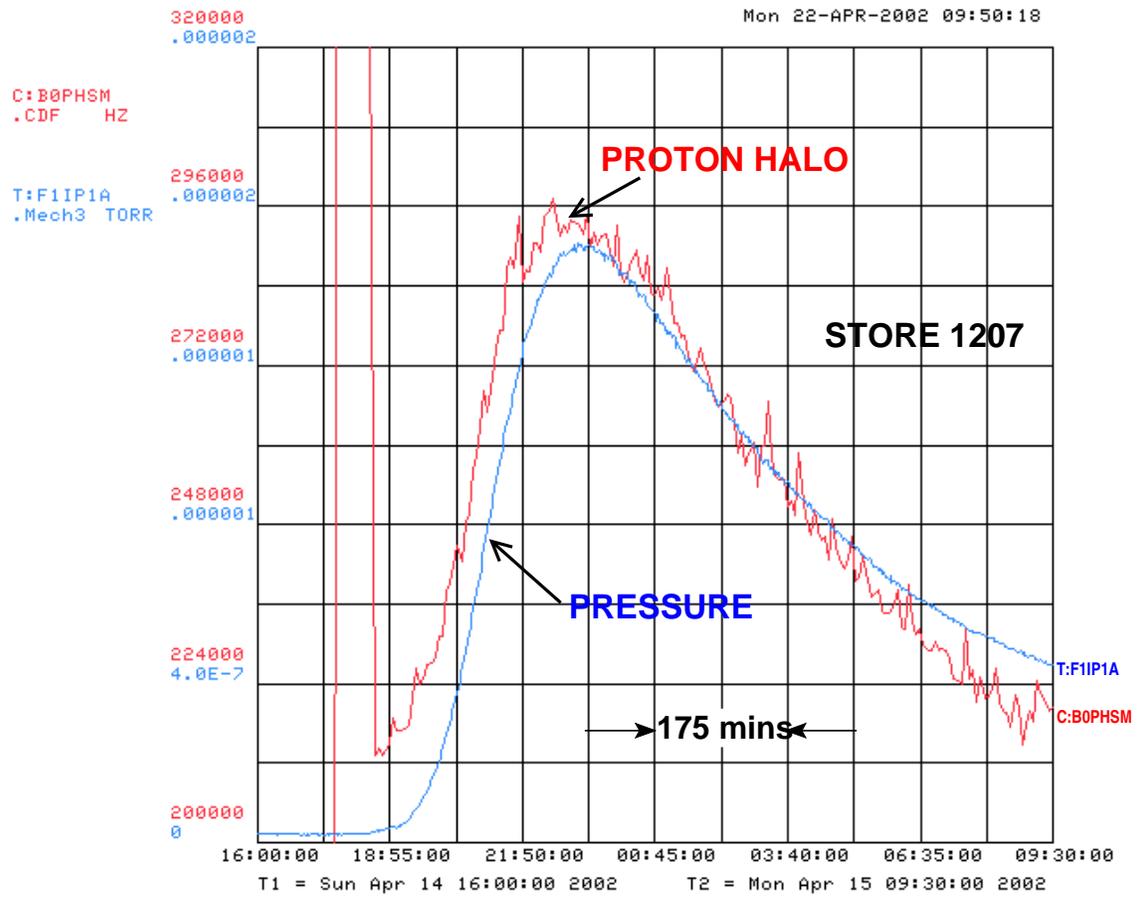} 
  \caption{\label{vacuum} 
            Total proton halo rate (C:B0PHSM) and the vacuum 
            measured in the Tevatron F11 sector (T:F1IPA) vs time 
            for store~1207.
            }
\end{center}
\end{figure}

\begin{figure}[ptb]
\begin{center}
  \epsfxsize=\figwidth in
  \epsffile{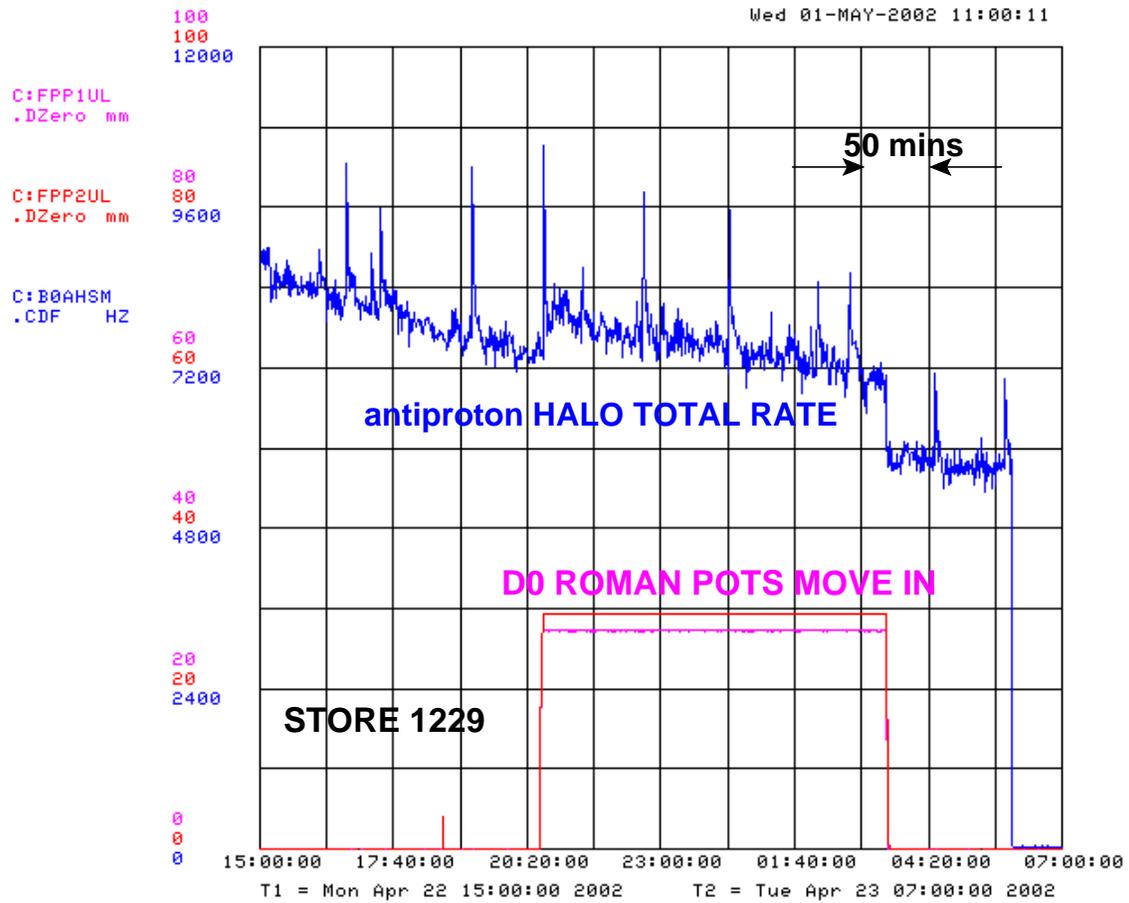}
  \caption{\label{romanpots}
            Total antiproton halo rate (C:B0AHSM) 
            and the D\O\ roman pot positions vs time during store~1229. 
            The antiproton halo rate increases as the D\O\ roman pots 
            are inserted.
            } 
\end{center}
\end{figure}

\begin{figure}[ptb]
\begin{center}
  \epsfxsize=\figwidth in
  \epsffile{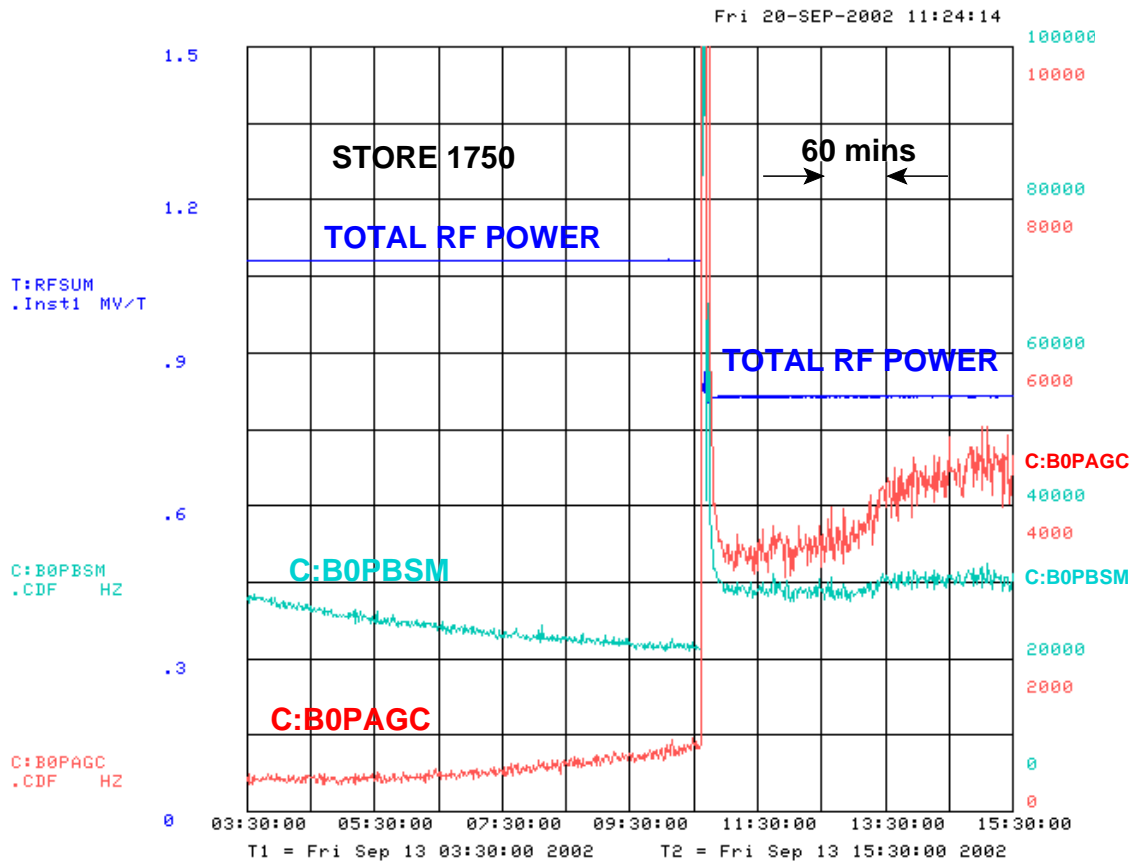}
  \caption{\label{rfcavity}  
            Total proton abort gap halo (C:B0PBSM) and  2/4 majority 
            coincidence (C:B0PAGC) rates vs time for a portion of 
            Tevatron store~1750. The power was lost for an RF cavity around 
            10:20AM on September~13th, 2002.
            }
\end{center}
\end{figure}

The halo monitors have also been observed to correlate well with 
features in the Tevatron accelerator. 
Listed here, are the sensitivity to changes in the beam tune, beam 
vacuum, RF power, Tevatron electron lens (TEL)~\cite{tel} failures 
and D\O\ Roman Pot \cite{d0romanpot} positions with respect to the beam 
line. Figures~\ref{pabort}--\ref{rfcavity} illustrate these correlations.
 Figure \ref{pabort} shows the rates for the abort gap monitors taken 
during store~1229. The halo in the abort gap monitors have a clear 
response to the TEL operational changes. The rates show 
high frequency features following the lens being turned off (0.0~mA). 
 As the TEL is turned back on, the rates increase abruptly with a 
subsequent decrease~\cite{rick}.  

Figure~\ref{vacuum} illustrates the correlation between the changes 
in the beam vacuum (T:F1IP1A) and proton halo (C:B0PHSM) rates during 
store~1207. The former variable is the vacuum ion gauge for the F11 
sector of the Tevatron. The vacuum problems, identified here, were 
subsequently addressed. The vacuum pressure improved from values just 
under $2\times10^{-6}$~Torr to well below $2\times10^{-8}$~Torr yielding 
about 30--40\% halo reduction in CDF Collision Hall. 

The monitors are also capable of detecting changes along the beam lines. 
Figure~\ref{romanpots} shows an abrupt increase in the antiproton halo 
monitor rates (C:B0AHSM) as a response to the D\O\ Roman Pot insertion 
during store~1229. The D\O\ detector is located at the D\O\ section of 
Tevatron, upstream of CDF in the antiproton direction. As of the writing of 
this article, we anticipate a controlled test with the pots to be performed. 

The abort gap halo rates correlate well with sudden losses within 
the RF structure in Tevatron. In such cases, more particles  
escape from the bunches and fill the abort gaps.  
Figure~\ref{rfcavity} includes that portion of Tevatron store~1750 at 
which the power to an RF cavity is lost.  Typically, the 
rates also continuously rise near the end of a store due to time 
dependent detoriation of the beam tune. This is reflected in the 
abort gap halo monitor (C:B0PAGC) rates.  

The halo monitors are in the regular monitoring list of the CDF 
experiment as the halo has an impact on silicon and muon detectors.  
Alarms are currently notifying experimenters on shift when beam 
conditions move out of tolerances.

\section{Conclusion}
We installed two counter arrays in the CDF collision hall to improve 
the monitoring of beam quality.  These counters are set up to monitor 
Tevatron halo and beam in the abort gaps with the proper choice of 
signal coincidences. The coincidences behave as expected under controlled 
conditions and new variables are installed in the Fermilab accelerator 
controls network (ACNET). Such monitors did not exist for Tevatron 
experiments during Run I. Data from these devices are shown to reproduce 
losses qualitatively and the devices are observed to be sensitive to 
effects not previously seen by existing monitors. The data also show 
that some beam effects are visible with an enhanced sensitivity when 
compared with existing detectors. The monitors are being used by the 
Fermilab accelerator experts in order to investigate the properties of the 
Tevatron beam halo as measured in the CDF collision hall. We believe 
these type of monitors also serve as examples for the next generation of 
hadron colliders for which the background and radiation conditions 
will be more severe.

\begin{ack}
We are very grateful to Dennis Nicklaus from the Fermilab Beams 
Division for software support and to Ron Moore (Fermilab, Beams Division) 
and Rick Vidal (Fermilab, Particle Physics Division-CDF) for pointing out various correlations.  
We would also like to thank Dervin Allen, Lew Morris, George Wyatt, 
Roberto Davila and Jamie Grado for their help during the installation 
of the monitors. M.K.U. acknowledges support under US Department of
Energy grant DE-FG02-91ER40684.
\end{ack}

\newpage

\begin{table}[tb]
 \begin{center}
 \caption{\label{scintillators}Operating parameters for scintillator 
          counters used in the CDF collision hall alcoves.}
 \vspace{2pt}
 \begin{tabular}{cccc}
  \multicolumn{1}{c}{~Counter~} & 
  \multicolumn{1}{c}{PMT Bias}  &
 \multicolumn{1}{c}{Attenuation} &
  \multicolumn{1}{c}{Ungated Dark Count} \\

  \multicolumn{1}{c}{ID}         &
  \multicolumn{1}{c}{Voltage (V)}     &
  \multicolumn{1}{c}{Length (cm)}  &
  \multicolumn{1}{c}{Rate (Hz)}          \\ \hline \hline 
   083 & 1875 & $294\pm 31$ 
        & 50 \\
   059 & 1637 & $241\pm 41$
        & 91 \\
   042 & 1888 & $238\pm 20$
        & 32 \\
   003 & 1785 & $255\pm 23$
        & 184 \\  \hline
   054 & 1668 & $282\pm 28$
        & 114 \\
   066 & 1495 & $229\pm 18$
        & 84 \\
   014 & 1843 & $260\pm 24$
        & 22  \\
   040 & 1786 & $255\pm 22$
        & 60 \\
 \end{tabular}
 \end{center}
\end{table}

\newpage
 \clearpage 

\begin{table}[tb]
 \begin{center}
 \caption{\label{nobeam} Measured and calculated beam-off rates 
          for the halo and abort gap monitors.} 
 \vspace{2pt}
 \begin{tabular}{cccccc}
  \multicolumn{2}{c}{Counter} &
  \multicolumn{2}{c}{Halo Rate (Hz)} & 
  \multicolumn{2}{c}{Abort Gap Rate (Hz)} \\
  \multicolumn{1}{c}{ID}   &
  \multicolumn{1}{c}{Device}   &
  \multicolumn{1}{c}{Meas.}    &
  \multicolumn{1}{c}{Calc.} &
  \multicolumn{1}{c}{Meas.}  &
  \multicolumn{1}{c}{Calc.} \\ \hline \hline
 083 &  B0A*N &  4$\pm$2   
              &  3$\pm$2       
              & 22$\pm$5   
              & 17$\pm$ 4     \\ 
 059 &  B0A*T &  5$\pm$2   
              &  5$\pm$2       
              & 36$\pm$6   
              & 32$\pm$ 6     \\ 
 042 &  B0A*S &  2$\pm$1   
              &  2$\pm$1       
              & 14$\pm$4   
              & 11$\pm$ 3     \\ 
 003 &  B0A*B & 11$\pm$3   
              & 11$\pm$3     
              & 78$\pm$9   
              & 64$\pm$ 8     \\ 
 054 &  B0P*N &  6$\pm$ 2  
              &  7$\pm$  3     
              & 45$\pm$7   
              & 40$\pm$ 6     \\ 
 066 &  B0P*T &  5$\pm$ 2  
              &  5$\pm$ 2     
              & 43$\pm$7   
              & 30$\pm$ 5     \\ 
 014 &  B0P*S &  1$\pm$ 1  
              &  1$\pm$1     
              &  8$\pm$ 3   
              &  8$\pm$  3     \\ 
 040 &  B0P*B &  3$\pm$ 2  
              &  4$\pm$ 2     
              & 21$\pm$5   
              & 21$\pm$ 5     \\ 
 \end{tabular}
 \end{center}
\end{table}

\clearpage
\pagebreak[4]
\clearpage
\clearpage
\clearpage
\clearpage
\clearpage
\clearpage
\clearpage
%
\clearpage
\clearpage
\vspace{15pt}

\end{document}